\documentclass[11pt,a4paper]{article} 

\usepackage{anysize}
\usepackage[format = hang]{caption}
\usepackage{amsmath,amsthm,verbatim,amssymb,amsfonts,amscd, graphicx}
\usepackage{graphics,natbib}
\usepackage{titlesec,footmisc}
\usepackage{hyperref} 

\usepackage{listings}
\usepackage{booktabs} 

\theoremstyle{plain}

\theoremstyle{definition}

\def\beq{\begin{eqnarray}}
\def\eeq{\end{eqnarray}}

\def\beqn{\begin{eqnarray*}}  
\def\eeqn{\end{eqnarray*}}

\titleformat{\section}{\normalfont\large\sc\centering}{\thesection}{1em}{}
\titleformat{\subsection}[runin]{\normalfont\large\bfseries}{\thesubsection}{1em}{}
\numberwithin{equation}{section} 
\renewenvironment{abstract}
               {\list{}{\rightmargin\leftmargin}%
                \item[\text{\hspace{10mm}\sc Abstract.}]\relax}
               {\endlist}

\begin{document}

\def\heute{March 2026}

\begingroup
\begin{centering} 
\Large{\bf Counting the uncounted: \\ How many were killed in Guatemala, 1978-1995?}
  \\[0.8em]
\large{\bf Nils Lid Hjort} \\[0.3em] 
\small {\sc Department of Mathematics, University of Oslo} \\[0.3em]
\small {\sc June 2026\footnote{invited paper, for an invited talk,
  at {\it The 40th International Workshop on Statistical Modelling},
  Oslo, June 28 to July 3, 2026}}\par 
\end{centering}
\endgroup



\begin{abstract}
\small{In various application domains, there is a certain
  `null cell', inside a multinomial setup, where observations
  are recorded for the other cells, but where one cannot
  count the number of occurrences for the null cell.
  I develop inference theory for assessing such unknown numbers,
  counting the uncounted, in situations where counts
  are available for the other cells, 
  via parametric modelling. The methods are used to
  estimate the number of persons killed in Guatemala
  during the {\it Genocidio guatemalteco} years 1978--1995.
  There are three carefully curated lists of
  killed people, where the information can be mapped to
  a Venn diagram with $2^3=8$ cells. Summing over the seven
  observed cells, $R=\hbox{47,803}$ killed individuals
  can be identified, but how big is $N_{0,0,0}$,
  and hence $N=N_{0,0,0}+R$?} 

\end{abstract}


\section{Who killed the bishop, and who killed 127,000 people?}
\label{section:hjort_section1}


\begin{figure}[!ht]\centering
\includegraphics[width=8cm]{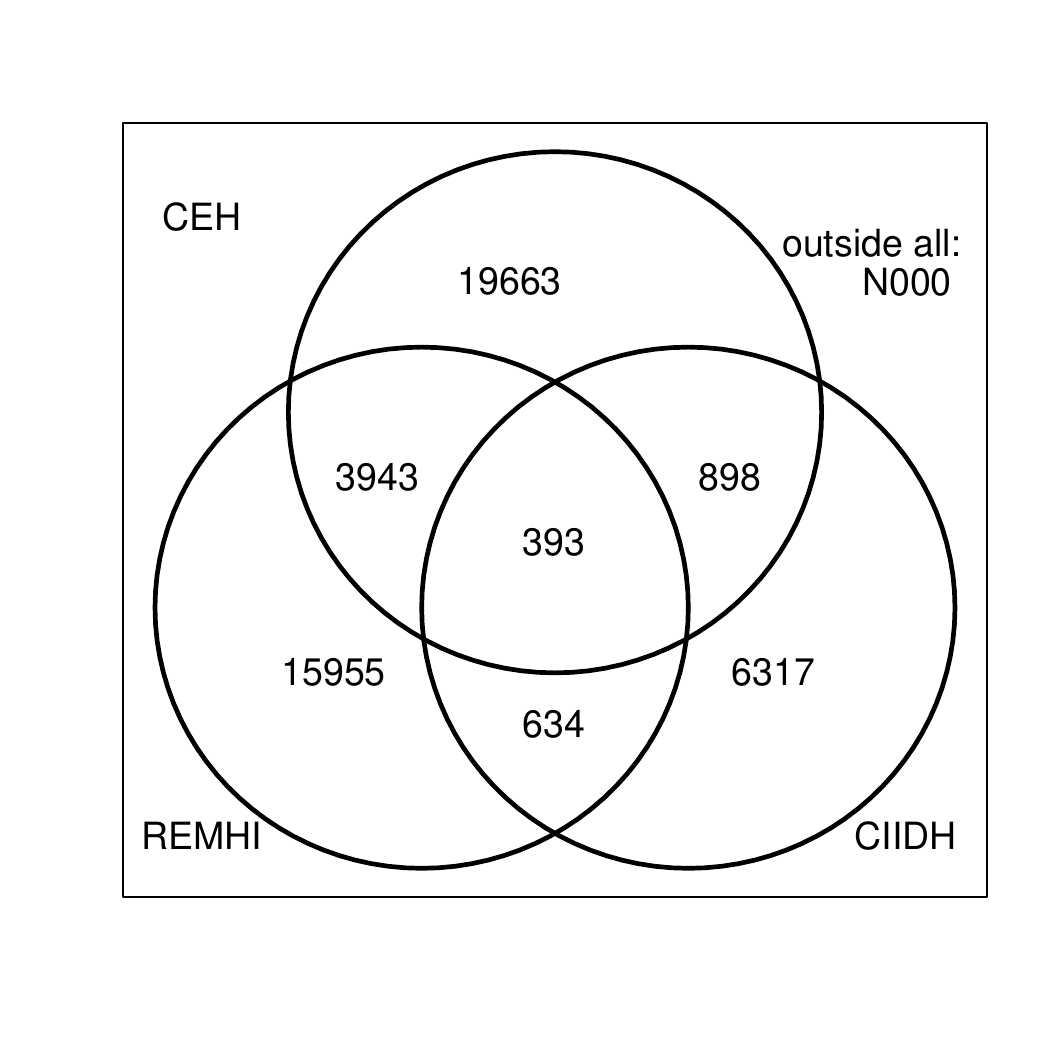} 
\caption{\label{figure:hjort_figure1}
Venn diagram for the number of people killed and accounted
for, for the three lists REMHI, CEH, CIIDH, in total
$R=\hbox{47,803}$ individuals. But how many
are not counted, in the N000 box?}
\end{figure}

Figure \ref{figure:hjort_figure1} is a dramatic one.
It pertains to people killed in Guatemala, during the tumultuous
years 1978--1995 of the {\it Genocidio guatemalteco},
or {\it Holocausto silencioso}, 
recorded via official lists. The three sources in question are
the Recovery of Historical Memory (REMHI),
Commission for Historical Clarification (CEH),
and the International Center for Human Rights Investigations (CIIDH),
with acronyms reflecting project names in Spanish.
The data, via careful scrutiny of lists from these
three organisations, can in Venn diagram terms be translated to
$N_{1,1,1} = \hbox{393}$,
$N_{1,1,0} = \hbox{3,943}$,
$N_{1,0,0} = \hbox{15,955}$, 
$N_{1,0,1} = \hbox{634}$,
$N_{0,1,1} = \hbox{898}$,
$N_{0,1,0} = \hbox{19,663}$, 
$N_{0,0,1} = \hbox{6,317}$.  
The total number of confirmed deads, captured via these three
lists, is
$R=\sum_{(i,j,k)\not=(0,0,0)} N_{i,j,k}=\hbox{47,803}$. 

For glimpses into these horror tales, and to the political,
sociological, racist, mass-psychological intricacies which
led to these turmoils and the systematic murdering,
carried out largely by state military and allied forces
during periods of U.S.~and CIA support for the Guatemalan regime,
check Ball (1999), Lum et al.~(2013), and Goldman's (2010) detailed account
{\it The Art of Political Murder: Who Killed Bishop Gerardi?}. 
As President B.~Clinton acknowledged, in March 1999, 
``support for military forces or intelligence units which engaged
in violent and widespread repression of the kind described
in the report was wrong, and the United States must not repeat that mistake''. 

The more mundane but nevertheless important challenge
worked with here is the statistical one, to estimate the full
number $N = N_{0,0,0}+\cdots+N_{1,1,1}$ of individuals killed,
hence also in the process the number $N_{0,0,0}=N-R$
of deads not captured on any of the three lists.

\section{Theory for multinomials with an unobserved cell}
\label{section:hjort_section2}

Suppose $(N_0,N_1,\ldots,N_k)$ is a multinomial vector,
corresponding to $N$ independent experiments,
associated with categories $0,1,\ldots,k$, with probabilities
$p_0,p_1,\ldots,p_k$. We observe $N_1,\ldots,N_k$,
hence their sum $R=N_1+\cdots+N_k$, but not $N_0$ for the zero-box,
falling outside the other categories.

We assume a smooth parametric model is being used for the probabilities,
with $p_j=p_j(\theta)$ for $j=0,1,\ldots,k$, with $\theta$
of dimension $r\le k-1$, admitting two continuous derivatives.
The joint probability distribution is
$$ f=\frac{N!}{N_0!\,N_1!\cdots N_k!} p_0(\theta)^{N_0} p_1(\theta)^{N_1} \cdots p_k(\theta)^{N_k}, $$
and the log-likelihood may be written
\begin{eqnarray*}
\ell(N,\theta)
=\log(N!)-\log((N-R)!)+(N-R)\log p_0(\theta) +\sum_{j=1}^k N_j\log p_j(\theta). 
\end{eqnarray*}
For the given model this also leads to the profiled log-likelihood
\begin{eqnarray*}
\ell_{{\rm prof}}(N)=\max\{\ell(N,\theta)\colon {\rm all\ }\theta\}
   =\ell(N,\widehat\theta_N). 
\end{eqnarray*} 
Maximising this function gives the maximum likelihood (ML)
estimators $(\widehat N,\widehat\theta)$. 

To reach clear and useful approximation statements for inference,
for growing $N$, assume that the model holds,
for some true $\theta_0$, and write $N_0$ for the true $N$.
The ML estimators
$(\widehat N,\widehat\theta)$ converge towards limits at
different rates, and it is fruitful to write $N=N_0\rho$
and $\widehat\rho=\widehat N/N_0$. This leads to the approximate
log-likelihood 
\begin{eqnarray*}
\ell(\rho,\theta)
&=&N_0\rho\log(N_0\rho)-(N_0\rho-R)\log(N_0\rho-R) \\
& &\qquad 
   +\,(N_0\rho-R)\log p_0(\theta)+\sum_{j=1}^k N_j\log p_j(\theta). 
\end{eqnarray*} 

From basic multinormal limits for multinomials, 
there is joint convergence
\beqn
N_0^{1/2}\{N_j/N-p_j(\theta_0)\}\rightarrow_d A_j,
\eeqn 
say, a zero-mean multinormal vector with 
${{\rm cov}}(A_j,A_\ell)=p_j(\theta_0)\{\delta_{j,\ell}-p_\ell(\theta_0)\}$;  
note that $A_0+\cdots+A_k=0$. With $\widetilde p_0=1-R/N_0$, we then have 
$N_0^{1/2}\{\widetilde p_0-p_0(\theta_0)\}\rightarrow_d A_0$. 

Letting $u_j(\theta)=\partial\log p_j(\theta)/\partial\theta$,
taking the derivative of $\sum_{j=0}^k p_j(\theta)=1$
leads to $\sum_{j=0}^k p_j(\theta)u_j(\theta)=0$. 
Below, write for crispness $p_j$ and $u_j$ for $p_j(\theta_0)$
and $u_j(\theta_0)$. Formulae for the first-order 
partial derivatives of the log-likelihood $\ell(\rho,\theta)$
can be put up, and when evaluated at $(N_0,\theta_0)$, these become 
\begin{eqnarray*}
U_0&=&N_0(-\log\widetilde p_0+\log p_0), \\
V_0&=&N_0\Bigl\{(\widetilde p_0-p_0) u_0
   +\sum_{j=1}^k (N_j/N_0-p_j)u_j\Bigr\}. 
\end{eqnarray*} 
A key quantity is $M=\sum_{j=0}^k p_ju_ju_j^T$. 
There is convergence in distribution 
$$\frac{1}{N_0^{1/2}} 
\begin{pmatrix}U_0 \\ V_0 \end{pmatrix} 
\rightarrow_d 
\begin{pmatrix}U \\ V \end{pmatrix}
= \begin{pmatrix}-(1/p_0)A_0 \\ \sum_{j=0}^k A_ju_j \end{pmatrix}
\sim{{\rm N}}_{r+1}(0,J),
$$
with
$$
J=\begin{pmatrix} (1-p_0)/p_0, &-u_0^T \\ 
   -u_0, &M \end{pmatrix}. 
$$
This is the Fisher information matrix for this reparametrised 
model, with parameters $(\rho,\theta)$, evaluated at $(1,\theta_0)$. 
We may verify that $J$ is also the limit 
in probability of $-N_0^{-1}H_0$, where $H_0$ is the $(r+1)\times(r+1)$
Hessian matrix of second order partial derivatives 
of $\ell(\rho,\theta)$, evaluated at $(1,\theta_0)$.

From this, we deduce first that
\begin{eqnarray}
N_0^{1/2}(\widehat N/N_0-1)=(\widehat N-N_0)/N_0^{1/2}
   \rightarrow_d {{\rm N}}(0,\tau^2),
\label{eq:limiting}
\end{eqnarray} 
where 
$$
\tau^2
=\frac{1}{(1-p_0)/p_0 - u_0^T M^{-1}u_0}
=\frac{p_0}{1-p_0 - p_0\delta},
\quad {\rm with\ }\delta=u_0^T M^{-1}u_0. 
$$
Though the main concern might be estimating $N_0$ and $N=N_0+R$,
we note also that
$N_0^{1/2}(\widehat\theta-\theta_0)\rightarrow_d {{\rm N}}_r(0,K)$, with
$$
K
=(M-(1-p_0)u_0u_0^T/p_0)^{-1}
=M^{-1} + dM^{-1}u_0u_0^T M^{-1},
$$
where we write
$$
d=r_0/(1-r_0\delta)=p_0/(1-p_0-p_0\delta)
\quad{\rm and}\quad r_0=p_0/(1-p_0).
$$
Interestingly, had $N_0$ been observed too,
the result would have been ${{\rm N}}_r(0,M^{-1})$, 
so $dM^{-1}u_0u_0^T M^{-1}$ is the extra amount of uncertainty
when estimating $\theta$, caused by $N_0$ not being counted.

The limit distribution result (\ref{eq:limiting}) is informative,
by giving the precision of the ML estimator,
and can also be used to form approximate confidence intervals.
It is easier, and often better, however, to use the associated
Wilks type chi-squared methodology. Indeed it can be shown,
as for classical parametric likelihood inference, that
\begin{eqnarray}
D(N_0)=2\{\ell_{{{\rm prof}},\max}-\ell_{{\rm prof}}(N_0)\}\rightarrow_d \chi^2_1. 
\label{eq:deviance}
\end{eqnarray} 
This leads to confidence intervals, e.g.~$\{N_0\colon D(N_0)\le 1.96^2\}$,
and a full confidence curve ${{\rm cc}}(N)=\Gamma_1(D(N_0))$, 
which we can read off directly from the profiled log-likelihood, 
in effect bypassing the perhaps cumbersome estimation of $\tau$
via the the $J$ matrix etc. 
For pertinent details, see Hjort and Stoltenberg (2026, Ch.~5)
and Schweder and Hjort (2016, Ch.~3--4). 

\section{The three-parameter independent lists model} 
\label{section:hjort_section3}

\begin{figure}[!ht]\centering
\includegraphics[width=8cm]{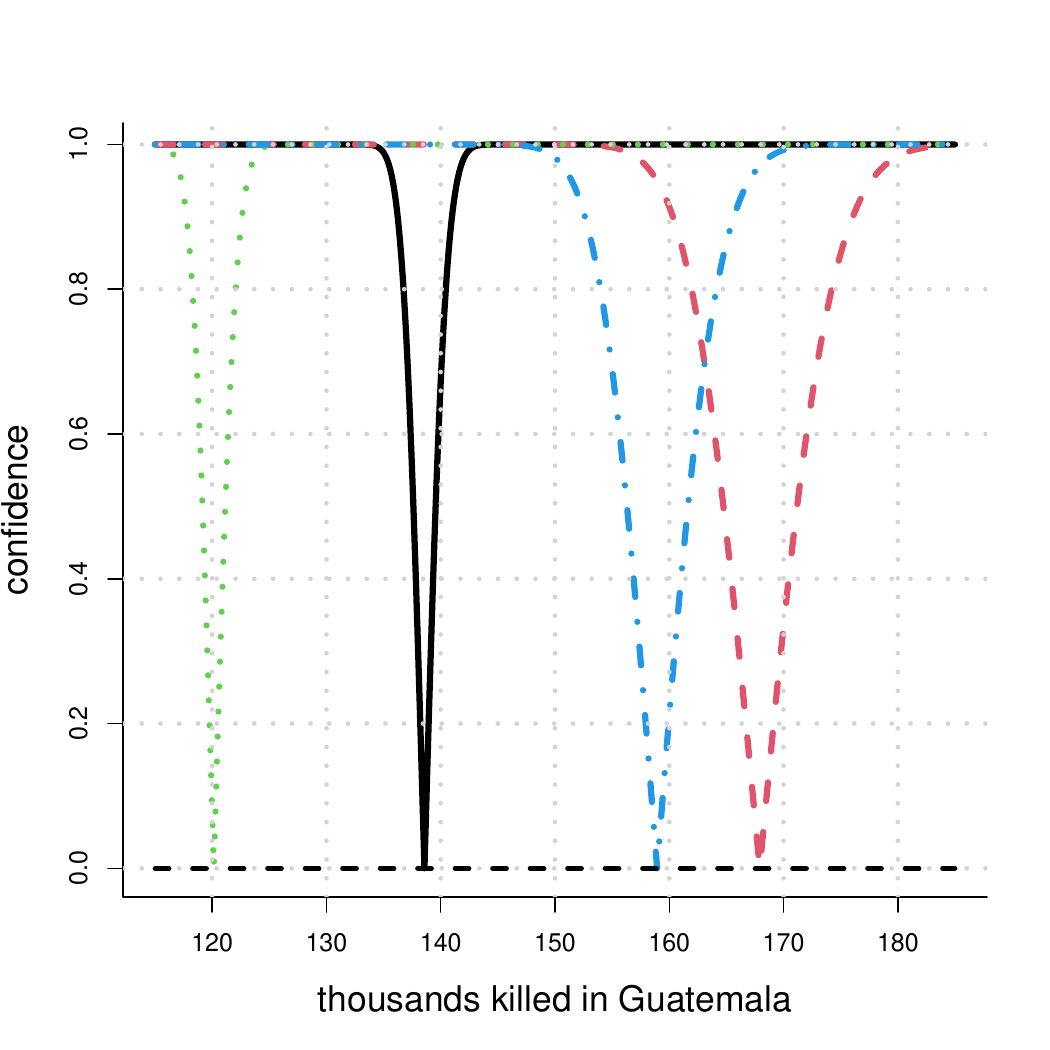}
\caption{\label{figure:hjort_figure2}
Confidence curves for $N$, the total number of people
killed in Guatemala 1978–1996, in thousands, based on
list independence, using all three sources (full black curve),
and using pairwise analyses. The best estimate,
under this assumption, is 138,576, with 95 percent interval 
$[\hbox{135,794},\hbox{141,453}]$.}
\end{figure}

A structurally simple and sometimes fully adequate model
for the cell probabilities is that of {\it list independence}.
The model postulates probabilities for a single person
to land in lists One, Two, Three to be $p,q,r$.
With independence, the eight cell probabilities then become
$p_{0,0,0}=(1-p)(1-q)(1-r)$, $p_{0,0,1}=(1-p)(1-q)r$, and so on,
up to $p_{1,1,1}=pqr$. The log-likelihood function becomes
\begin{eqnarray*}
\ell(N,p,q,r)&=&\log (N!)-\log((N-R)!)
   +A\log p+(N-A)\log(1-p) \\
& &\!\!\!\!\!\!   
   +B\log q+(N-B)\log(1-q)
   +C\log r+(N-C)\log(1-r),
\end{eqnarray*} 
in terms of $A=N_{1,\cdot,\cdot}=N_{1,0,0}+N_{1,0,1}+N_{1,1,0}+N_{1,1,1}$,
and similarly $B=N_{\cdot,1,\cdot}$, $C=N_{\cdot,\cdot,1}$.
This is easily maximised over $p,q,r$ for each given $N$,
leading to
$$ \ell_{{\rm prof}}(N)=\log(N!)-\log((N-R)!)
   +N\{H(\widehat p_N)+H(\widehat q_N)+H(\widehat r_N)\}, $$
with $\widehat p_N=A/N$, $\widehat q_N=B/N$, $\widehat r_N=C/N$,
involving the function $H(u)=u\log u+(1-u)\log(1-u)$. 

Using the general theory of Section \ref{section:hjort_section2}, 
related to the precise limiting distribution of (\ref{eq:limiting}), 
some work shows first that the $M$ matrix is diagonal with
elements $1/\{p(1-p)\}$, $1/\{q(1-q)\}$, $1/\{r(1-r)\}$, and that 
$$ \delta=\psi_{0,0,0}^T M^{-1}\psi_{0,0,0} = p/(1-p)+q/(1-q)+r/(1-r). $$
From this we conclude that
$(\widehat N-N)/N^{1/2}\rightarrow_d{{\rm N}}(0,\tau^2)$, with
$$ \tau^2=\frac{ (1-p)(1-q)(1-r) }{1-(1-p)(1-q)(1-r)(1+\delta)}
   =\frac{ (1-p)(1-q)(1-r) }{pq + pr + qr - 2pqr}. $$
This result is a 3-list extension of known results
in the capture-recapture literature for the 2-list case. 
It is seen that the variance increases drastically
with smaller inclusion probabilities $p,q,r$. 

The simpler case of only two independent sources can indeed
be treated similarly, then involving the profiled
log-likelihood function
$$
\ell_{{\rm prof}}(N)=\log(N!)-\log((N-R)!)
   + N\{H(\widehat p_N)+H(\widehat q_N)\}
$$
and $\widehat p_N=A/N$, $\widehat q_N=B/N$, with $A=N_{1,\cdot}$ 
and $B=N_{\cdot,1}$. One may show that the ML
estimator is large-sample equivalent to and in fact
very close to the first published estimator
for such capture-recapture problems,
the Petersen (1896) estimator $N^*=AB/N_{1,1}$.
It was used by him to estimate the number of young plaice
immigrating to the Danish Limfjord from the German Sea;
see also Hjort and Petersen (1905). 

Partly bypassing this ${{\rm N}}(0,\tau^2)$ result, we may 
use the deviance function (\ref{eq:deviance}) and the associated 
Wilks type result we to compute ${{\rm cc}}_3(N)$,
the confidence curve for $N$ using this 3-parameter model.
Figure \ref{figure:hjort_figure2} displays this three-sources
based ${{\rm cc}}_3(N)$, alongside those obtained using
only two sources at the time. The point estimate is
$\widehat N_3=\hbox{138,576}$, with 95 percent
confidence interval $[\hbox{135,794},\hbox{141,453}]$.
Point estimates for the inclusion probabilities $(p,q,r)$
are $(0.151,0.180,0.059)$, i.e.~rather small, implying
high chances of a killed individual not being detected;
this outside probability $(1-p)(1-q)(1-r)$ is estimated
at 0.655. This spells statistical difficulty for
assessing the zero cell $N_{0,0,0}$, and indeed also
nonrobustness for the $\widehat N_3$. 


\section{Better models, with four and five parameters} 
\label{section:hjort_section4}

\begin{figure}[!ht]\centering
\includegraphics[width=8cm]{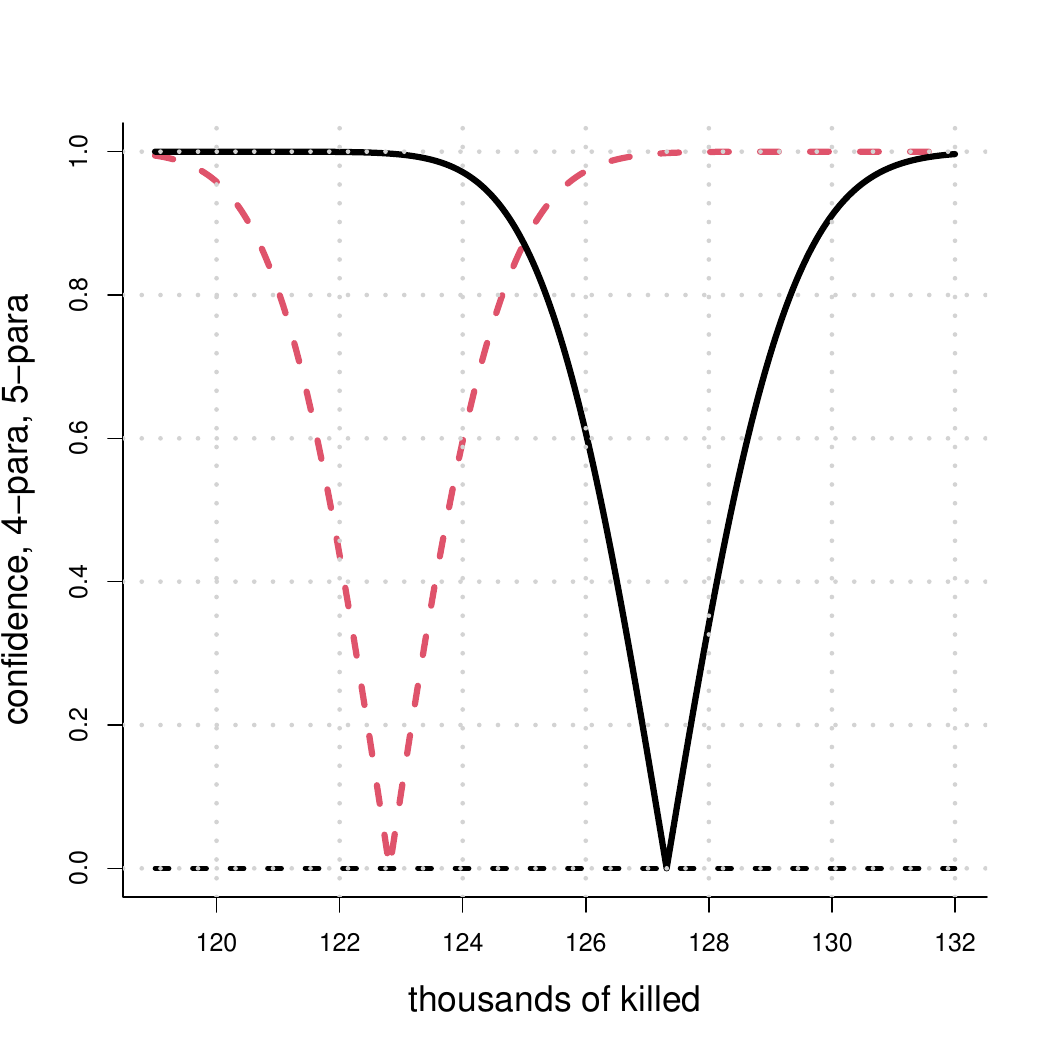}
\caption{\label{hjort_figure3}
Confidence curves for $N$, the number of persons killed in Guatemala,
in thousands, using the 4-parametric (red, dashed)
and 5-parametric (black, full) models.
Note that the $N$ scale used here is shorter than for
Figure \ref{figure:hjort_figure2}. The point estimates
are 122,812 and 127,314, with 95 percent intervals
$[\hbox{120,100}, \hbox{125,634}]$ and $[\hbox{124,341}, \hbox{130,415}]$.} 
\end{figure}

%
%

There are several reasons indicating that the list independence
assumption is not holding up for the Guatemala killings, however,
as pointed to in Ball (1999) and Lum et al. (2013). That is also born
out statistically, as we shall see now, when we construct first
a better 4-parameter model and then an even better and
eventually satisfactory 5-parameter model.
Figure \ref{figure:hjort_figure2}, with the difference between
the pairwise estimates, is also evidence against independence. 

Consider first the four-parameter model 
\begin{eqnarray*}
p_{0,0,0}&=&(1-p)(1-q)(1-r)/s, \\
p_{0,0,1}&=&(1-p)(1-q)r\,\gamma/s, \\
p_{0,1,0}&=&(1-p)q(1-r)/s, \\
p_{0,1,1}&=&(1-p)qr/s, \\
p_{1,0,0}&=&p(1-q)(1-r)/s, \\
p_{1,0,1}&=&p(1-q)r/s, \\ 
p_{1,1,0}&=&pq(1-r)/s, \\
p_{1,1,1}&=&pqr/s,  
\end{eqnarray*} 
where the $\gamma$ is a parameter associated with cell 001, 
modifying independence in that direction, and $s$ is the factor
required to give sum 1 over the eight cells. 
We can then numerically maximise the log-likelihood
$\ell(N,p,q,r,\gamma)$, and learn that the Karl Pearson type statistic 
$$
K=\sum_{i,j,k} \frac{ (N_{i,j,k}
     - \widehat N\widehat p_{i,j,k})^2}{\widehat N\widehat p_{i,j,k}}
$$
is much smaller than for the simpler three-parameter 
independence model; here $\widehat N_{0,0,0}=\widehat N-R$
is used for $N_{0,0,0}$.
Checking the numerics we find that the $K$ is reduced
from 487.05 with the 3-parametric model to 220.31
for the 4-parametric model;
also, the log-likelihood max is increased with as much as 156.22. 
The ML estimates are $\widehat N_3=\hbox{138,576}$
and $\widehat N_4=\hbox{122,812}$. 

It turns out that an even better model is as follows, with
five parameters. It uses push or modification factors
$\gamma_1$ and $\gamma_2$, placed for cells 001 and 111, 
\begin{eqnarray*}
p_{0,0,0}&=&(1-p)(1-q)(1-r)/s, \\
p_{0,0,1}&=&(1-p)(1-q)r\,\gamma_1/s, \\
p_{0,1,0}&=&(1-p)q(1-r)/s, \\
p_{0,1,1}&=&(1-p)qr/s, \\
p_{1,0,0}&=&p(1-q)(1-r)/s, \\
p_{1,0,1}&=&p(1-q)r/s, \\ 
p_{1,1,0}&=&pq(1-r)/s, \\
p_{1,1,1}&=&pqr\,\gamma_2/s,  
\end{eqnarray*} 
and again with $s$ the required factor securing these
eight probabilities having sum 1.
There are actually 28 such models, placing $\gamma_1$ and $\gamma_2$
different places, so to speak, but the one above, involving
such pushes away from independence for cells 001 and 111,
is seen to be the clearly best one, in terms of log-likelihood increase
and smaller value of the Pearson $K$.
The push parameters $(\gamma_1,\gamma_2)$
are estimated at $(1.85,2.31)$; in particular, there is a higher
chance for a killed person than merely $pqr$ of landing
in each of the three lists. In yet other words,
if you have been detected once, the chance is bigger that you
will be detected by one or both of the others. 

The informative summary Table 1 gives the expected
or predicted numbers $\widehat E_{i,j,k}=\widehat N\widehat p_{i,j,k}$,
for the three models, as exp3, exp4, exp5, and then the
Pearson residuals $P_{i,j,k}=(N_{i,j,k}-E_{i,j,k})/E_{i,j,k}^{1/2}$,
as pear3, pear4, pear5. For the null cell, I use
$\widehat N_{0,0,0}=\widehat N\widehat p_{0,0,0}$.
As we see, there are clear discrepancies between data
and the 3- and 4-parameter models, whereas the 5-parameter
model passes muster and is judged to deliver the most precise
estimates. From Figure \ref{hjort_figure3},
an estimated $\widehat N_5=\hbox{127,314}$ individuals were
killed, with confidence interval $[\hbox{120,100}, \hbox{125,634}]$. 

\begin{table}[!ht]\centering
\caption{Estimates of $N_{0,0,0}$, with expected numbers $E_{i,j,k}$
for the other seven categories, along with Pearson residuals
$P_{i,j,k}=(N_{i,j,k} - E_{i,j,k})/E_{i,j,k}^{1/2}$, for the 3-, 4-, 5-parameter
models. The Pearson type statistic $K=\sum_{i,j,k} P_{i,j,k}^2$
takes the values 487.05, 220.31, 6.91 for the 3-, 4-, 5-parameter models,
the latter being the clear and adequate winner.}
\medskip

\begin{small}
\begin{verbatim}
        obs        exp3      exp4      exp5    pear3   pear4   pear5
  000         90773.343 75009.236 79511.180   -0.001  -0.001  -0.001
  001   6317   5740.222  6316.999  6317.008    7.612   0.000  -0.001
  010  19663  19880.337 19606.796 19712.903   -1.541   0.401  -0.355
  011    898   1257.170   954.011   847.908  -10.129  -1.813   1.720
  100  15955  16144.568 15819.072 15904.699   -1.491   1.080   0.398
  101    634   1020.932   769.711   684.106  -12.109  -4.891  -1.915
  110   3943   3535.834  4134.975  3943.191    6.847  -2.985  -0.003
  111    393    223.595   201.196   393.002   11.329  13.522  -0.001
                                             487.052 220.308   6.914 
\end{verbatim} 
\end{small} 

\end{table} 

\section{Concluding remarks} 
\label{section:hjort_section5}

Some of the comments below point to rhe need for more
research. 

\smallskip
{\it Srebrenica 1995.}
A partly similar but statistically somewhat
easier tale is that of counting the killed in Srebrenica, 1995.
There are two lists, not three, but the inclusion probabilities
$p,q$ are higher than for the Guatemala case.
See Hjort and Stoltenberg (2026, Story \#65) for details,
with point estimate $\widehat N=\hbox{7,543}$ for the number killed. 

\smallskip 
{\it FIC.}
Bartolucci and Lupparelli (2008) have developed
a Focused Information Criterion, in the style of Claeskens and Hjort (2008),
applicable to such multinomial data with missing observations.
There is a need to generalise this further, building on
methods from this paper. A broader FIC can be built,
e.g.~for the purpose of sorting through and ranking
the 28 5-parameter candidate models of the type used in Section~3,
via (i) putting up a fixed wide model, assumed to contain
an approximation to the true data-generating model,
and (ii) establishing and estimating biases, variances,
mean squared errors, for the resulting $N$ estimators.
This requires careful extensions of the material of
Section 2, to setups where the parametric models used
are outside the correct data-generating process.
Ongoing work in such directions will be reported on elsewhere. 

\smallskip 
{\it Bayes.} 
Methods developed and worked with in this paper have been
frequentist in spirit, but parallel Bayesian methods can
be constructed too, via appropriate priors for either
$N$ or $N_{0,0,0}$ themselves, or for the parameters
of the models used. One may e.g.~use information from
Ball (1999) to elicit priors for the inclusion probabilities
$p,q,r$, both for the simpler 3-parameter model of Section 2,
and for the more complex ones of Section~3. 

\smallskip
{\it Further methodological advances.}
There is a need to develop and refine models and methods
for assessing null cells in multinomial or Poisson setups,
as this paper can be seen as an invitation to.
See in this connection also
the recent PhD dissertation Zuit (2024),
and models worked with in Hjort (2026) for negatively
correlated binomials and Poissons. 

\smallskip
{\it Other uses of the methodology.}
Examples abound in various application domains, where
one needs to `count the uncounted'. How many people are
cheating on their taxes; how many illegal immigrants
are there; how many words did Shakespeare know (apart
from those he actually used); how many unseen sharks
are there in a certain region. And, sadly, one may continue
to hunt, as does Patrick Ball and the
{\it Human Rights Data Analysis Group}, for the number
of killed people, in wars and in conflict zones. 


%
%


{\bf Acknowledgments.}
I am grateful for many fruitful discussions
  on themes related to this article with C\'eline Cunen.
  Thanks are also due to the participants of the
  {\it Stability and Change} programme, led by N.L.~Hjort and H.~Hegre, 
  at the Centre for Advanced Study,
  the Academy of Science and Letters, Oslo,
  August 2022 to June 2023, with political scientists,
  war and conflict scholars, and statisticians.
  The theme of this particular paper, `estimation of those
  not accounted for', was among those studied during that
  programme, from many perspectives, and applied for several
  partly similar dramatic historical events of war and conflict.  
  In particular, Patrick Ball contributed both detailed information
  and general wisdom pertaining to the Guatemala statistics
  and to similar studies.

\section*{References}

\begin{small}
\parindent0pt
\parskip3pt

Ball, P. (1999). 
  Making the Case: Investigating Large Scale Human Rights Violations
  Using Information Systems and Data Analysis.
  {\it American Academy for the Advancement of Science}, Washington

Bartolucci, F.~and Lupparelli, M. (2008).
Focused Information Criterion for capture-recapture models
for closed populations. {\it Scandinavian Journal of Statistics},
{\bf 9}, 658--664.

Claeskens, G.~and Hjort, N.L. (2008).
  {\it Model Selection and Model Averaging.}
  Cambridge: Cambridge University Press. 

Goldman, F. (2010).
  {\it The Art of Political Murder: Who Killed Bishop Gerardi?}
  New York: Atlantic Books. 

Hjort, J.~and Petersen, C.G.J. (1905).
Short review of the results of the international fisheries
investigation, volume 3.
{\it Conseil Permanent International Pour l’Exploration de la Mer,
Rapport et Proces-Verbaux des Reunions}, K{\o}benhavn.

Hjort, N.L. (2026).
Modelling pairs of Poissons and binomials
with negative correlation.
Statistical Research Report, Department of Mathematics,
University of Oslo.

Hjort, N.L.~and Stoltenberg, E.Aa. (2026).
  {\it Statistical Inference: 600 Exercises, 100 Stories.}
  Cambridge: Cambridge University Press (in process).

Lum, K., Price, M. E., and Banks, D. (2013).
  Applications of multiple systems estimation in human
  rights research. {\it American Statistician}, {\bf 24}, 191--200.

Petersen, C.G.J. (1896).
The yearly immigration of young plaice into the Limfjord from the German Sea.
{\it Report of the Danish Biological Station}, {\bf 6}, 5--84.

Schweder, T.~and Hjort, N.L. (2016).
  {\it Confidence, Likelihood, Probability: Statistical Inference
  via Confidence Distributions.}
  Cambridge: Cambridge University Press. 

Zuit, D.B. (2024). 
  {\it Counting the Uncounted: Methodological Extensions
  in Multiple Systems Estimation.}
  PhD Dissertation, Universiteit Utrecht. 

\end{small}

\end{document}